\documentclass[reprint,
twocolumn,
english,
aps,
prb,
superscriptaddress,
bibnotes,
amsmath,
amssymb,
floatfix,
longbibliography,
citeautoscript,
]{revtex4-2}

\usepackage[table]{xcolor}
\usepackage[utf8]{inputenc}
\usepackage{array}
\usepackage[T1]{fontenc}
\usepackage[defaultcolor=red, final]{changes}
\usepackage[colorlinks=true, allcolors=blue]{hyperref}
\usepackage{graphicx, dcolumn, siunitx, orcidlink}
\usepackage{threeparttable}
\usepackage{makecell}
\definecolor{mygray}{gray}{0.95}

\graphicspath{{figures/}}

\newcommand{\chitwo}{$\chi^{(2)}$}
\newcommand{\chitwoeff}{$\chi^{(2)}_\text{eff}$}
\newcommand{\chithree}{$\chi^{(3)}$}
\newcommand{\SiN}[0]{Si$_3$N$_4$}
\newcommand{\wtwow}{$\omega$-2$\omega~$}

\begin{document}

\title{Reconfigurable Resonant Multimode Nonlinear Coupling\\ for UV-to-infrared Frequency Generation}

\author{Samantha Sbarra\orcidlink{0000-0002-0893-927X}}
\thanks{These authors contributed equally to this work.}
\affiliation{Photonic Systems Laboratory (PHOSL), STI-IEM, {\'E}cole Polytechnique F{\'e}d{\'e}rale de Lausanne,  CH-1015 Lausanne, Switzerland}

\author{Ji Zhou\orcidlink{0000-0001-8044-4426}}
\thanks{These authors contributed equally to this work.}
\affiliation{Photonic Systems Laboratory (PHOSL), STI-IEM, {\'E}cole Polytechnique F{\'e}d{\'e}rale de Lausanne, CH-1015 Lausanne, Switzerland}

\author{Boris Zabelich\orcidlink{0000-0002-8542-7542}}
\affiliation{Photonic Systems Laboratory (PHOSL), STI-IEM, {\'E}cole Polytechnique F{\'e}d{\'e}rale de Lausanne,  CH-1015 Lausanne, Switzerland}

\author{Marco Clementi\orcidlink{0000-0003-4034-4337}}
\affiliation{Photonic Systems Laboratory (PHOSL), STI-IEM, {\'E}cole Polytechnique F{\'e}d{\'e}rale de Lausanne,  CH-1015 Lausanne, Switzerland}
\affiliation{Present address: Dipartimento di Fisica “A. Volta”, Università di Pavia, 27100 Pavia, Italy}

\author{Christian Lafforgue\orcidlink{0009-0003-2027-9168}}
\affiliation{Photonic Systems Laboratory (PHOSL), STI-IEM, {\'E}cole Polytechnique F{\'e}d{\'e}rale de Lausanne,  CH-1015 Lausanne, Switzerland}

\author{Ozan Yakar\orcidlink{0000-0003-1357-8920}}
\affiliation{Photonic Systems Laboratory (PHOSL), STI-IEM, {\'E}cole Polytechnique F{\'e}d{\'e}rale de Lausanne,  CH-1015 Lausanne, Switzerland}

\author{Junqiu Liu\orcidlink{0000-0003-2405-6028}}
\affiliation{Laboratory of Photonics and Quantum Measurements (LPQM), SB-IPHYS, {\'E}cole Polytechnique F{\'e}d{\'e}rale de Lausanne,  CH-1015 Lausanne, Switzerland}
\affiliation{Present address: International Quantum Academy and Shenzhen Futian SUSTech Institute for Quantum Technology and Engineering, Shenzhen 518048, China}
\affiliation{Present address: Hefei National Laboratory, University of Science and Technology of China, Hefei 230088, China}

\author{Tobias J. Kippenberg\orcidlink{0000-0002-3408-886X}}
\affiliation{Laboratory of Photonics and Quantum Measurements (LPQM), SB-IPHYS, {\'E}cole Polytechnique F{\'e}d{\'e}rale de Lausanne,  CH-1015 Lausanne, Switzerland}

\author{Camille-Sophie Br\`es\orcidlink{0000-0003-2804-1675}}
\email{\textcolor{magenta}{camille.bres@epfl.ch}}
\affiliation{Photonic Systems Laboratory (PHOSL), STI-IEM, {\'E}cole Polytechnique F{\'e}d{\'e}rale de Lausanne,  CH-1015 Lausanne, Switzerland}

\date{\today}
% Abstract: max 200 words
\begin{abstract}
%\begin{linenumbers}
\noindent 
%TC:ignore
On-chip coherent visible and near-infrared (NIR) light generation has broad applications in metrology, bio-sensing, and quantum information. High-Q microresonators are ideal candidates for generating light across such broad wavelength ranges via efficient second- (\chitwo) and third-order (\chithree) nonlinear optical processes. However, harnessing these diverse nonlinearities simultaneously in a single microresonator remains elusive yet highly attractive both fundamentally and technologically. Here, we demonstrate coherent light generation from the ultraviolet to NIR in a silicon nitride microresonator pumped by a single continuous-wave telecom laser. This broad frequency generation arises from the interplay of \chitwo\ and \chithree\ nonlinear processes. A cascade of nonlinear processes, including harmonic generation and optical parametric oscillation (OPO), is initiated by the photo-induced second harmonic generation enabled by all-optical poling. The dynamic reconfigurability of this \chitwo\ nonlinearity enables access to different transverse spatial modes at the second harmonic, enabling highly tunable OPO processes triggered by hybrid modal phase matching conditions and yielding  milliwatt-level NIR light. This work sheds new insights into the fundamental physics of cooperative nonlinear multimode interactions in resonant systems and provides a versatile approach for reconfigurable OPOs, highlighting their potential to generate light at wavelengths beyond the reach of photonic integrated lasers.
%TC:endignore
\end{abstract}

\maketitle

% \linenumbers

\noindent 
High-Q microring resonators exploit tight light confinement to enhance light–matter interactions and precisely control dispersion, enabling low-power nonlinear processes with unprecedented efficiency at visible and near-infrared (NIR) wavelengths
\cite{Gaeta2019Photonicchipbased, Lu2024Emergingintegratedlaser, Dutt2024Nonlinearquantumphotonics}. Leveraging  second-order ($\chi^{(2)}$) and third-order ($\chi^{(3)}$) nonlinearities, nonlinear frequency conversion processes such as harmonic generation (HG), sum-frequency generation (SFG), four-wave mixing (FWM), and optical parametric oscillation (OPO), can provide generation of coherent light spanning from the visible to the NIR band, using continuous-wave (CW) pumps at technologically mature wavelengths.
Among these processes, HG and SFG yield spectral coverage strictly determined by the pump frequency, whereas FWM and OPO can be designed with tailored dispersion to access spectral components flexibly distant from the pump laser \cite{Sayson2019Octavespanningtunable, Fujii2017a, Sayson2019Octavespanningtunable, Fujii2019, Lu2020chipopticalparametric, Tang2020, Domeneguetti2021Parametricsidebandgeneration, Lu2019Milliwattthresholdvisible}. 
Combining \chitwo and \chithree\ nonlinear processes can generate either monochromatic or broadband light across widely separated spectral regions, linking them together in a coherent fashion.
Despite this potential, practical applications have so far been limited mainly to supercontinuum generation with $\chi^{(2)}$ and $\chi^{(3)}$  spectral extension \cite{Wu2024Visibleultravioletfrequency, Ludwig2024Ultravioletastronomicalspectrograph, Bres2023Supercontinuumintegratedphotonics}, and comb self-referencing \cite{Okawachi2018carrierenvelope,Hickstein2019Selforganizednonlinear, Okawachi2020Chipbasedself,Nitiss2020broadbandquasi} in single-pass waveguides. 
Recently, harnessing cooperative \chitwo\ and \chithree\ nonlinearities in high-Q microresonators has attracted significant attention.
Such cascaded interactions have been successfully implemented for the synthesis of optical nonlinearities \cite{Wang2022Syntheticfivewave, Cui2022situcontroleffective, Cui2025ReconfigurableSelfPhase}, comb generation, and spectral translation in multiple integrated photonic platforms, including aluminum nitride \cite{Jung2014GreenredIR,Guo2018efficient,Liu2018Generationmultiplevisible}, lithium niobate \cite{He2019selfstarting}, gallium phosphide \cite{Wilson2020Integratedgalliumphosphide}, silicon carbide \cite{Wang2022solitonformation}, and silicon nitride (\SiN) \cite{Miller2014,Clementi2025ultrabroadband,mehrabad2025multi,Xue2017Secondharmonicassisted,Hu2022Photoinducedcascaded,Li2023highcoherence}.

Among these, \SiN\ is an established platform \cite{Morin2021CMOSfoundrybased, Chauhan2022Ultralowloss, CoratoZanarella2023Widelytunablenarrow, Smith2023SiNfoundryplatform,Liu2021Highyield}, featuring high power capability, a wide transparency window ranging from the mid-infrared to UV, and excellent \chithree\ performance.
Despite lacking an intrinsic \chitwo\ due to its amorphous nature, recent works have shown that \SiN\ can be endowed with an effective \chitwo\ (\chitwoeff) through all-optical poling (AOP) \cite{Billat2017Largesecond, Hickstein2019Selforganizednonlinear, Nitiss2020FormationRulesDynamics, Nitiss2022Opticallyreconfigurablequasi, Yakar2022GeneralizedCoherentPhotogalvanic}.
Remarkably, this photo-induced nonlinearity can self-adapt both its periodicity and transverse profile to form a self-organized \chitwoeff\ grating, naturally satisfying the quasi-phase-matching (QPM) condition between the pump and any resonant second harmonic (SH) mode, thereby enabling efficient \cite{Lu2021efficientphotoinduced}, reconfigurable \cite{Nitiss2022Opticallyreconfigurablequasi, Zhou2025Selforganizedspatiotemporal, Yanagimoto2025}, high-yield \cite{mehrabad2025multi}, highly tunable and broadband SHG \cite{Clementi2025ultrabroadband}. 
Despite these attractive properties, the interplay between such \chitwoeff\ and \chithree\ nonlinearities in \SiN\ microresonators has so far been limited to Kerr comb initiation under normal dispersion \cite{Xue2017Secondharmonicassisted, Hu2022Photoinducedcascaded} and spectral translation via SHG/SFG \cite{Miller2014, mehrabad2025multi, Clementi2025ultrabroadband, wang2025integrated}.
The high SH power and modal versatility enabled by AOP open opportunities towards further functionalities that still remain largely underexplored.

\begin{figure*}[t]
\centering
\includegraphics[width=1\linewidth]{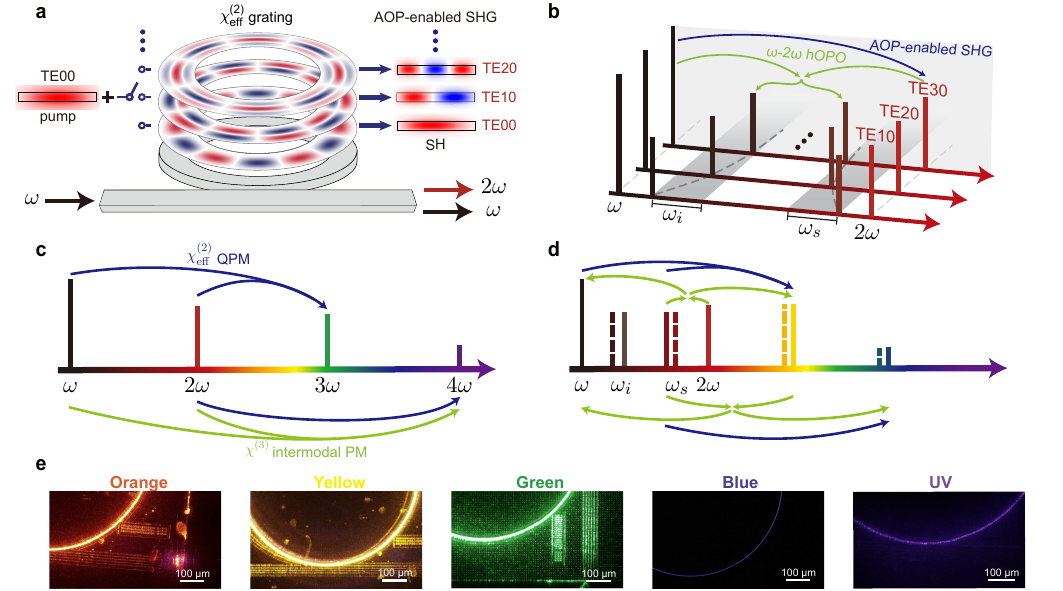}
    \caption{\textbf{UV-to-NIR coherent light generation in a high-Q \SiN\ microresonator from a single telecom pump.}
    \textbf{a}. Illustration of optically reconfigurable second-harmonic generation in a \SiN\ microring resonator: the photo-induced nonlinearity from all-optical poling self-organizes into a QPM \chitwoeff\ grating whose spatial profile is determined by the nonlinear interference between the TE00 pump and distinct SH modes (TE00, TE10, TE20...) enabling efficient SHG across various SH modes. 
    \textbf{b}. A hybrid optical parametric oscillation with two harmonically spaced pumps (\wtwow hOPO) can generate signal ($\omega_{\mathrm{s}}$) and idler ($\omega_{\mathrm{i}}$) spectrally located between the pump frequencies. Dynamical switching between different SH modes alters the phase-matching condition for the \wtwow hOPO process through re-configuration of the nonlinear grating, providing wide tunability of the signal and idler frequencies across the NIR range.
    \textbf{c}. Simultaneous second-, third-, and fourth-harmonic generation achieved via intermodal phase-matched \chithree\ (green arrows) or \chitwoeff-mediated (blue arrows) processes.
    \textbf{d}. Visible light generation beyond pump harmonic processes triggered by the \wtwow hOPO-generated signal and idler waves through further nonlinear interactions such as stimulated four-wave mixing, sum-frequency generation, and second-harmonic generation yielding yellow, orange, and blue colors. 
    \textbf{e}. Top-view images of the visible light scattered from the microresonator.
}
\label{fig1}
\end{figure*}

In this work, we demonstrate the generation of two-octave-spanning light from ultraviolet (UV) to NIR range in a high-Q \SiN\ microresonator by leveraging the cooperation of photo-induced \chitwoeff\ and \chithree\ nonlinearities. Driven by a single monochromatic CW telecom pump, the multi-color generation initially onsets from an efficient AOP-enabled SHG.
This seeds cascaded high-harmonic generation of green and UV light, while simultaneously initiating a dual-pump \chithree-OPO process that produces tunable, milliwatt-level signal and idler outputs between two harmonically spaced pumps. 
Intriguingly, further cascaded processes involving the OPO signal/idler fields complete the access to a wide range of colors in the visible spectrum. 

The strength of this scheme lies in its ability to selectively and effectively excite distinct modal orders at the SH, achieved through a self-adaptive redistribution of space charges in the waveguide cross-section. This reconfigurability is used here for the first time as a fundamental degree of freedom to drastically control the phase coupling of the \chithree-OPO process that follows, increasing the tunability beyond what  is otherwise possible by adjusting the pump wavelength, temperature, and/or sample geometry, as seen so far in OPO driven by non-reconfigurable pumps. Our work exploits resonant multimode nonlinear coupling between the pump and its SH to generate tunable CW coherent light across a wide spectral range, beyond the reach of integrated lasers and individual nonlinear processes. This capability arises from the synergy of reconfigurable SHG, widely tunable \chithree-OPO, and further nonlinear cascaded processes all exploiting resonant multimode nonlinear coupling.

\begin{figure*}[t]
\centering
\includegraphics[width=1\linewidth]{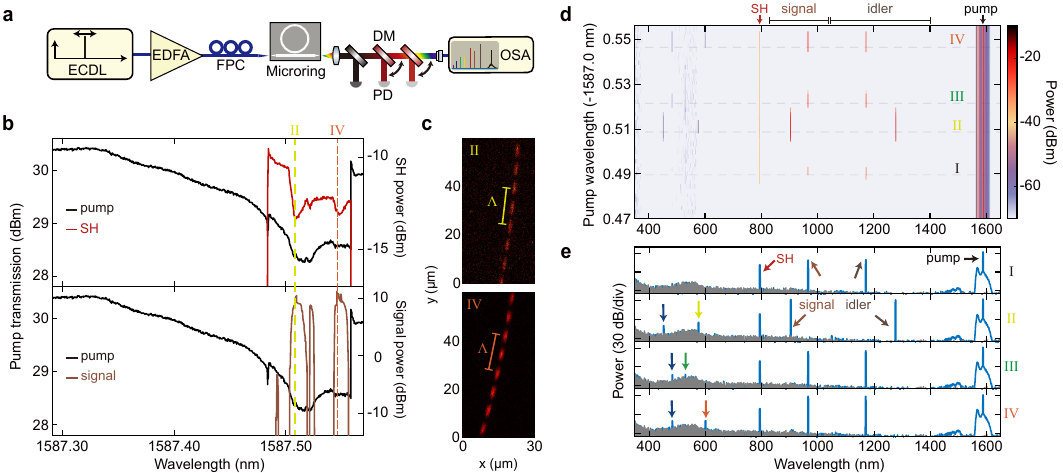}
\caption{
    \textbf{Experimental investigation of AOP-enabled SHG and cascaded nonlinear processes.} 
    \textbf{a}. Schematic of the experimental setup. The second and third dichroic mirrors (DMs) are removable to allow for either signal and SH power detection or broadband UV-to-NIR spectra acquisition. ECDL, external-cavity diode laser; EDFA, erbium-doped fiber amplifier; FPC, fiber polarization controller; PD, photodetector; OSA, optical spectrum analyzer.    
    \textbf{b}. On-chip pump transmission, SH power, and signal (800-1100 nm) power during a wavelength sweep across the pump resonance at 1,587.3~nm. 
    \textbf{c}. Two-photon microscopy images of the nonlinear \chitwoeff\ gratings along the ring circumference. The microresonator was poled at pump wavelengths corresponding to spectral positions II and IV in panel \textbf{b}. Both gratings feature a period of $\Lambda=15.5$ µm, confirming the generation of SH in the TE40 mode.
    \textbf{d}. UV-to-NIR spectral map obtained by tuning the pump wavelength across the same resonance shown in \textbf{b}.
    \textbf{e}. Slices of the spectral map in \textbf{d} at pump wavelengths of 1,587.489, 1,587.508, 1,587.521, and 1,587.546~nm (positions I to IV). Two distinct signal/idler pairs are observed at approximately 904.4/1,275.7 nm and 965.2/1,171.5 nm. The orange (599.6 nm), yellow (575.2 nm), and blue (452.2, 482.6, 482.8 nm) spectral lines are indicated by corresponding colored arrows. They however have poor extraction and out-coupling efficiency to the OSA.
    }
\label{fig2}
\end{figure*}

\section*{Results}
\noindent \textbf{\normalsize Principle of multimode nonlinear coupling}\\
Fig.~\ref{fig1}a and b illustrate, respectively, the AOP-enabled SHG and cascaded \chithree-OPO process in a \SiN\ microring resonator. Hereafter, we refer to this fundamental- SH driven OPO process as \wtwow hybrid OPO (hOPO), in reference to the octave separation between the two driving fields, and the hybrid mode nature of the interacting waves reminiscent of previous works \cite{Zhou2022HybridModeFamily, Perez2023HighperformanceKerr, Ng2023Widebandmultimodeoptical}.
The \chithree-hOPOs differ from single-mode OPOs in three main aspects. They feature high output power since they are pumped in a highly-normal dispersion regime that suppresses parasitic processes even under strong pumping, such as close-band FWM \cite{Kippenberg2004, Herr2012universal} and comb cluster formation \cite{Matsko2016Clusteredfrequencycomb, Sayson2018Originsclusteredfrequency}; they show robustness against geometric and temperature variations as they rely on modal effective refractive indices, rather than on higher-order dispersion \cite{Sayson2017Widelytunableoptical, Sayson2019Octavespanningtunable}; however, this also means that they have reduced tunability \cite{Zhou2022HybridModeFamily}.

Typical dual pump OPOs necessitate external coupling of both pump fields into the microresonator. By generating one of the pump fields internally through AOP, our approach allows flexible modal control via external excitation by only the fundamental pump mode. As shown in Fig.~\ref{fig1}b, the photo-induced \chitwoeff\ grating self-adapts its azimuthal periodicity and transverse profile to satisfy QPM between the TE00 pump and the corresponding resonant SH mode. Although the second pump ($2\omega$) is fixed as the SH of the input pump ($\omega$), the unlocked SH modal degree of freedom substantially reshapes the phase-matching landscape of the hOPO process, allowing exceptionally broad tunability of the signal ($\omega_{\mathrm{s}}$) and idler ($\omega_{\mathrm{i}}$) frequencies across the NIR range, as depicted in Fig.~\ref{fig1}b.

Extended wavelength conversion arises from higher-harmonic generation of the pump, reaching the green  ($3\omega$) and UV ($4\omega$) ranges. These cascaded \chitwoeff\ and/or $\chi^{(3)}$ processes, are enabled by the high SH power and the system's multimode versatility. Visible light generation is further enriched by cascaded interactions driven by the \wtwow hOPO signal and idler, as evidenced by top-view scattering images in Fig.~\ref{fig1}e. Orange/yellow emission emerges from stimulated FWM (sFWM) or SFG, while additional combinations can yield blue light (Fig. ~\ref{fig1}d), all consistent with energy conservation. An overview of all potential processes that can contribute to visible light generation is presented in Supplementary Note I. 

\begin{figure*}[bt]
\centering
\includegraphics[width=1\linewidth]{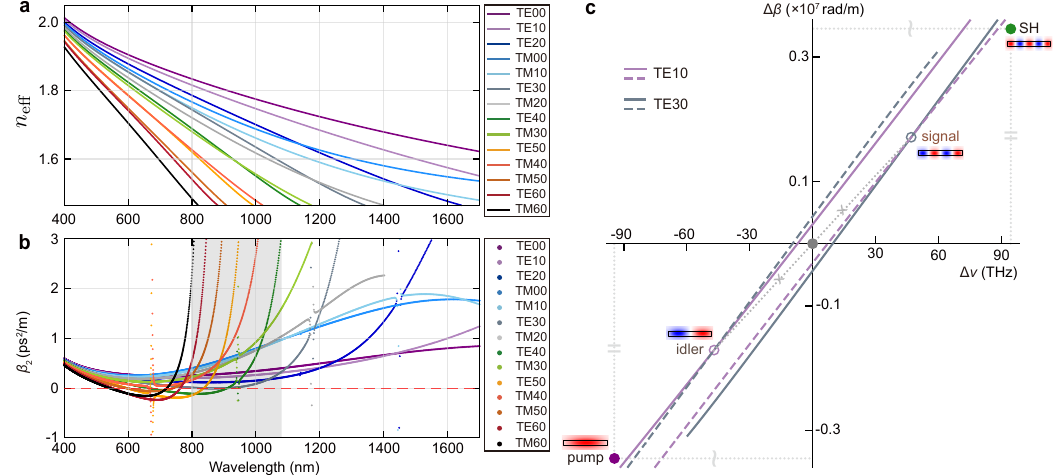}
\caption{
    \textbf{Simulation analysis of dispersion and phase-matching for the} \wtwow \textbf{hOPO.}
    \textbf{a}. Effective refractive indices ($n_\text{eff}$) for TE and TM modes up to the seventh order. Avoided mode crossing and mode hybridization occur at the intersections of specific TE and TM modes.  
    \textbf{b}. Group velocity dispersion ($\beta_2$) for TE and TM modes. Most modes exhibit normal dispersion in the NIR range, while anomalous dispersion is confined to the TE30, TE40, and TE50 modes within the SH and signal (gray-shaded area) bands. Avoided mode-crossings cause strong local perturbation of $\beta_2$.
    \textbf{c}. Graphical identification of signal and idler wavelengths for the \wtwow hOPO, driven by the TE00 pump at 1587.5~nm (violet dot) and the corresponding TE40 SH (green dot). The dispersion curves $\beta (\nu)$ are transformed to $\Delta \beta (\Delta \nu)$ (solid lines) using the relations $\Delta\beta = \beta - (\beta_\mathrm{pump} + \beta_\mathrm{SH})/2$ and $\Delta\nu = \nu - (\nu_\mathrm{pump} + \nu_\mathrm{SH})/2$. This sets the midpoint (gray dot) $((\nu_\mathrm{pump} + \nu_\mathrm{SH})/2, (\beta_\mathrm{pump} + \beta_\mathrm{SH})/2)$ as the origin. The dashed lines represent the mirror-symmetric counterparts of the solid ones, given by $-\Delta\beta(-\Delta\nu)$. Energy and momentum conservation restrict the possible signal/idler pairs to residing at the centrosymmetric intersections of these solid and dashed curves. The identified pair consists of a TE30 signal (dark gray circle) at 330.2 THz (907.9 nm) and a TE10 idler (light violet circle) at 236.3 THz (1,268.7 nm).
    }
\label{fig3}
\end{figure*}

\vspace{0.3cm}
\noindent \textbf{\normalsize Experimental investigation}\\
\noindent The experimental setup is illustrated in Fig.~\ref{fig2}a and detailed in Methods. 
The device under study is a silica-embedded microring resonator with a free spectral range of approximately 50 GHz.
The ring and bus waveguides have a cross-section of 2.6 µm $\times$ 0.3 µm  (width$\times$height). 

The fundamental TE00 mode near 1587.3 nm, is overcoupled with intrinsic and loaded quality factors of 13.7 and 2.3 million, respectively (see Supplementary Note II). 
In the top panel of Fig.~\ref{fig2}b, the on-chip pump transmission for this specific resonance measured, is shown in black, as the laser wavelength is swept from the blue to red side of the resonance. The input power is approximately 30.4 dBm, and the resonance exhibits the typical triangular profile owing to thermal and Kerr effects. The red trace shows the detected SH at the output of the chip. When the wavelength exceeds 1587.48 nm, SHG begins and persists until the end of the thermal triangle, resulting in a bandwidth of approximately 80 pm (9.5 GHz), consistent with prior observations \cite{Nitiss2022Opticallyreconfigurablequasi, mehrabad2025multi, zhou2025broadbandspectralmapping}. Such large bandwidth is a result of the dynamical optical reconfigurability of both the amplitude and phase of the photo-induced \chitwoeff\ nonlinearity, which significantly relaxes the doubly resonant condition for three-wave mixing processes \cite{zhou2025broadbandspectralmapping}.

When the intracavity pump and SH power exceed the OPO threshold and multiply resonant condition is satisfied, a signal/idler pair is generated, due to energy and momentum conservation.
In the bottom panel of Fig.~\ref{fig2}b, the optical power collected in the 800-1100 nm range reveals four distinct peaks, corresponding to the presence of \wtwow hOPO signals. We used two-photon microscopy to image the inscribed \chitwo\ grating responsible for the SHG at different detuning (Fig.~\ref{fig2}c).
In both cases, the azimuthal grating period $\Lambda$ was 15.5 µm, which unequivocally identifies, for these specific experimental conditions, the SH mode as TE40 from simulations (see Supplementary Note III).
The corresponding 2 nm-resolution spectral map from UV to telecom range is plotted in Fig.~\ref{fig2}d.
The map reveals that, for this specific resonance and pumping condition, two distinct signal/idler pairs can be generated during the laser sweep.
The first (hOPO1) and second (hOPO2) pairs appear at 965.2/1,171.4~nm and 904.4/1,276.0~nm, respectively. We observe several occurrences of hOPO1, which we attribute to resonance shifts of the pump, SH, signal and idler modes, owing to their different thermal shifts. As a result, the relative alignment of the cavity resonances evolves (as also seen in the pump transmission and SH power trace), leading to the observed dynamic hopping. The relaxed multiple resonance condition associated with photo-induced \chitwo\ still allows the process to be maintained efficiently.

Besides the NIR light, we observe the generation of yellow, blue, green, and orange light, as detailed in map slices presented in Fig.~\ref{fig2}e. These spectral components are generated by cascaded nonlinear processes involving the \wtwow hOPO fields, as illustrated in Fig.~\ref{fig1}c and d. 

\vspace{0.3cm}
\noindent \textbf{\normalsize Numerical analysis}\\
\noindent In standard hOPO processes, the frequency tuning ratio between the signal/idler and the input pump is typically close to 1:1 \cite{Zhou2022HybridModeFamily}, provided that the order of the participating modes remains unchanged. This is inconsistent with the wide spectral distance between hOPO1 and hOPO2 shown in Fig.~\ref{fig2}d.
Since the TE00 pump and TE40 SH remain fixed, this suggests that the modal order of the signal and/or idler varies during pump tuning.
To verify this hypothesis, we calculate the phase-matching condition for this FWM-based process, starting from simulated effective refractive indices $n_{\text{eff}}$ for supported TE and TM modes (Fig.~\ref{fig3}a). 
Fig.~\ref{fig3}b presents the corresponding group velocity dispersion $\beta_2$. It reveals that the telecom pump in the TE00 mode is in the highly normal dispersion regime, favoring the desired \chitwo$\Rightarrow$\chithree\ cascade over the reverse order. % which would otherwise onset from close-band FWM and/or degenerate OPO.
Within the hOPO signal band (gray-shaded region), anomalous dispersion is experienced by the TE30, TE40, and TE50 modes, 
where a parametrically driven Kerr comb could theoretically emerge \cite{ Moille2024Parametricallydrivenpure, weng2025hyperparametricsolitonsnondegenerateoptical}. We also predict avoided-crossing points between some specific TE and TM modes, mode hybridization induces local perturbations of both $n_{\text{eff}}$ and $\beta_2$. 

The FWM-based hOPO process must respect the fundamental criteria of energy and momentum conservation, which respectively require $\nu_{\text{pump}} + \nu_{\text{SH}} = \nu_{\text{signal}} + \nu_{\text{idler}}$ and $\beta_{\text{pump}}+\beta_{\text{SH}}=\beta_{\text{signal}}+\beta_{\text{idler}}$, with $\nu$ and $\beta$ denoting the optical frequencies and mode-propagation constants. 
These two conditions can be combined as 
\begin{equation}
		\Delta \beta _{\mathrm{signal}} (\Delta \nu _{\mathrm{signal}})=-\Delta \beta _{\mathrm{idler}} ( -\Delta \nu _{\mathrm{idler}})\\
\label{Energy Phase conservation}
\end{equation}
where $\Delta\nu_{\text{signal, idler}} =\nu_{\text{signal, idler}}-(\nu_{\text{pump}}+\nu_{\text{SH}})/2$ and $\Delta \beta_{\text{signal, idler}}=\beta_{\text{signal, idler}}-(\beta_{\text{pump}}+\beta_{\text{SH}})/2$.
This reformulation reveals that the allowed signal and idler are symmetric about the frequency and momentum midpoint between pump and SH.

With the midpoint set as the origin, Fig.~\ref{fig3}c plots the translated dispersion curves $\Delta\beta(\Delta\nu)$ for the representative TE10 and TE30 modes (solid lines), along with the TE40 SH (green dot) and TE00 pump (violet dot), reproducing the experimental scenario of hOPO2 from Fig.~\ref{fig2}. On the same plot, the dashed lines correspond to the centrally symmetric transformation of the solid curves with respect to the origin, i.e., $-\Delta\beta(-\Delta\nu)$. 
Following equation (\ref{Energy Phase conservation}), finding the allowed signal/idler pairs means locating the intersection points between the solid and dashed lines.
The signal (gray circle) appears in the first quadrant on the TE30 mode at 907.9 nm, while its paired idler (light violet circle) is found on the TE10 mode at 1268.7 nm in the third quadrant, with a small discrepancy of 1.3 THz with respect to the experiments. 
Extending this graphical approach to all modes, two further combinations are theoretically predicted for this specific case: the TE00/TE30 pair at 979.7/1150.8 nm and the TE20/TE20 pair at 1013.2/1107.9 nm (see Supplementary Note IV for the corresponding graphs). 
The former corresponds to hOPO1 from Fig.~\ref{fig2}, with a slight wavelength discrepancy arising from the TE30–TM20 mode hybridization around the idler and not included in our simulations.
While this theoretical approach predicts all possible \wtwow hOPO signal/idler pairs given one pump and SH mode combination, some may not reach the OPO threshold, and hence not be experimentally observed (as is the case for the TE20/TE20 combination). In addition, competition between nonlinear processes make hOPO pair switching more favorable than their coexistence, as seen in Fig.~\ref{fig2}d.
Nevertheless, different combinations can be accessed by fine tuning the pump wavelength and chip temperature, facilitated by the large number of transverse and longitudinal modes.

\begin{figure}[htbp]
\centering
\includegraphics[width=1\linewidth]{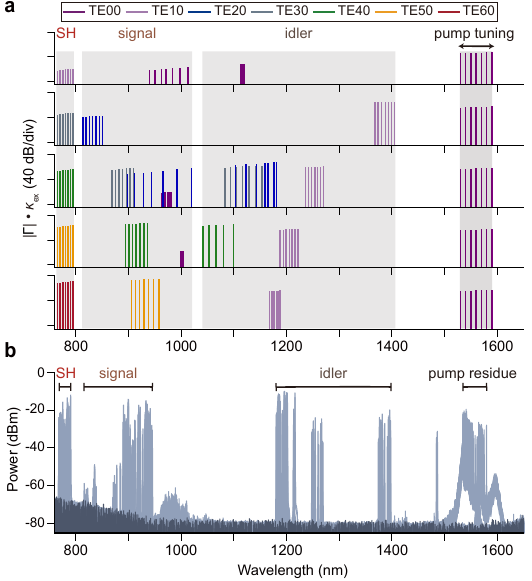}
\caption{
    \textbf{Coarse and fine tuning of OPO signal and idler wavelengths in the NIR range.} 
    \textbf{a}. Simulated \wtwow hOPO spectra with different pump wavelengths and SH modes.
    The pump wavelength is tuned over 60 nm range, while the SH mode is changed from TE00 to TE60. No \wtwow hOPO pair is found for SH in the TE00 and TE20 modes. The amplitude of the spectral lines are defined by the product of the modal overlap factor $\Gamma$ and the coupling coefficient to the bus waveguide $\kappa_\mathrm{ex}$. The excitation of TE40 SH enables up to three different combinations of modes, two of which were observed experimentally. 
   \textbf{b}. Overlaid experimental spectra recorded while tuning the pump wavelength (1535-1580 nm) and chip temperature (20-80$^{\circ}\mathrm{C}$). The focus of the output achromatic lens is always optimized for the SH coupling.
    }
\label{fig4}
\end{figure}

\noindent \textbf{\normalsize Wide and fine tuning of \wtwow hOPO}\\
\noindent Another tuning knob is to change the SH mode, which will greatly influence the hOPO phase-matching conditions.
For a comprehensive theoretical investigation, we extended the graphical method presented in Fig.~\ref{fig3}c to other TE00 pump wavelengths and SH modes, with the results shown in Fig.~\ref{fig4}a. 
We numerically swept the pump wavelength from 1530 nm to 1590 nm with 10 nm steps and present the resulting signal/idler pairs on distinct rows corresponding to each possible excited SH mode. 
The \wtwow hOPO interacting fields are shown at their spectral position, color-coded by their modal order. We defined the amplitude as the product of the mode overlap integral $\Gamma$ \cite{Lin2008proposalhighlytunable} and the coupling rate $\kappa_{\text{ex}}$ between the ring and bus waveguides (see Supplementary Note II).
Provided that the threshold power can be reached with sufficiently large $\Gamma$, this product scales with the expected output power of the generated spectral components.
Except for TE00 and TE20 SH mode combination whose dispersion does not allow any \wtwow hOPO process (hence not presented), Fig. \ref{fig4}a highlights the wide tuning of signal/idler wavelengths in the NIR range that can be achieved by switching the SH mode, together with the expected 1:1 fine tuning via pump wavelength adjustment.
For the identified \wtwow hOPO pairs, the idler primarily resides in the TE10 mode, whereas the signal is distributed among higher-order modes, facilitating their extraction from the microresonator.
Interestingly, even for optical modes that are nominally symmetry-mismatched, their large spectral distance can relax the symmetry constraints, resulting in a non-zero mode overlap.

We compared the numerical simulations with the experimentally collected NIR spectra obtained from changing both pump-wavelength and temperature. 
Experimentally, wide and fine tunings are intertwined due to the intricate dynamics of SH-mode competition. Furthermore, identification of the SH mode would require extensive TPM imaging for each acquired spectrum, and is not feasible.
For simplicity, we hence overlay all collected experimental spectra in Fig.~\ref{fig4}b.
We observe that, while dispersion in the off-chip coupling to the OSA prevents a direct comparison of peak amplitudes with theory, the overall spectral coverage shows good agreement with simulations. 
In the 950 - 1180 nm spectral window, the significantly reduced mode overlap ($\Gamma$ < 4$\%$) renders these hOPO processes inefficient and experimentally
inaccessible.

We estimated a maximum SHG conversion efficiency ($\eta_{\mathrm{SHG}}$) of 18$\%$/W (measured in the absence of any cascaded processes). This value is lower than those reported in Refs. \cite{Clementi2023achipscale, Li2023highcoherence} for devices with comparable Q factors, owing to our use of a higher pump power, but allows to reach high internal powers of the SH necessary for cascaded processes. 
The signal/idler power reaches milliwatt level, deduced from comparison with the SH coupling loss, corresponding to a conversion efficiency $\eta_{\mathrm{hOPO}} = (\omega_{\mathrm{pump}}/\omega_{\mathrm{signal, idler}}) \cdot (P_{\mathrm{signal, idler}}/P_{\mathrm{pump}})$ \cite{Stone2022a} of a few percent.
Experimentally, we were able to observe the \wtwow hOPO at input pump power down to 28 dBm, which could be reduced by improving the $\eta_{\mathrm{SHG}}$ in a high-finesse microresonator \cite{Lu2021efficientphotoinduced}.

\begin{figure}[h]
\centering
\includegraphics[width=1\linewidth]{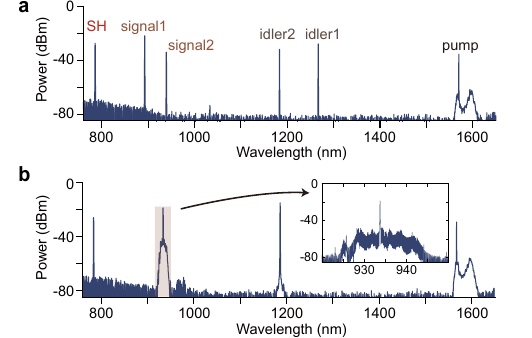}
\caption{
    \textbf{Other cascaded nonlinear processes in the \wtwow hOPO.} 
    \textbf{a}. Coexistence of two hOPO pairs for the pump/SH/signal/idler wavelengths at 1570.2/785.1/892.0/1266.6 nm and 1570.2/785.1/938.8/1182.9 nm, with the modes inferred as TE00/TE40/TE30/TE10 and TE00/TE60/TE50/TE10, respectively.
    \textbf{b}. Observation of a \wtwow hOPO followed by a MI comb at the signal band. The MI comb has a span of 20 nm. The pump/SH/signal/idler modes are TE00/TE50/TE40/TE10 with corresponding wavelengths of 1567.1/783.5/933.4/1186.1 nm.
    }
\label{fig5}
\end{figure}

\vspace{0.3cm}
\noindent \textbf{\normalsize Additional hOPO dynamics}\\
\noindent In addition to tuning dynamics of \wtwow hOPO, the multimode nonlinear coupling gives rise to several other intriguing phenomena.
Among them, we present in Fig.~\ref{fig5}a the coexistence of multiple signal/idler pairs originating from distinct SH modes, and in Fig.~\ref{fig5}b Kerr comb generation in the signal band. From simulations, we identified the signal mode as TE40 which experiences anomalous dispersion, allowing for the generation of a modulation instability (MI) comb. Still, the control of the comb generation is complex as the power and detuning of the comb pump are coupled through the hOPO process and further investigation are needed. In the Supplementary Note V, we report further hOPO dynamics.

\begin{table*}[htbp]
\renewcommand{\arraystretch}{1.3}
  \begin{threeparttable}
    \caption{Comparison of visible-to-NIR multicolor light generation via \chitwo\ and/or \chithree\ processes in state-of-the-art planar microresonator platforms.}
    \label{table1}
    \begin{center}
      \footnotesize\centering
      \begin{tabular}
        { 
        p{2cm}<{\centering} 
        p{5cm}<{\centering} 
        p{4cm}<{\centering} 
        p{3cm}<{\centering} 
        p{3.2cm}<{\centering} }
        \hline
        \textbf{\makecell[c]{Platform}} &
        \textbf{\makecell[c]{Nonlinear processes}} &
        \textbf{\makecell[c]{Wavelength access (nm)}} &
        \textbf{\makecell[c]{Pump \\ wavelength (nm)}} &
        \textbf{\makecell[c]{Max on-chip \\ power (mW)}} \\
        \hline

        LiNbO$_3$ \cite{KimContinuousWaveLaser} &
        SHG, \chitwo-SFG&
        387.94, 517.27, 775.88 &    % pump is in between 1040-1065 nm
        1551.75 &
        % 1 for 775.88 nm 
        n.a., n.a., 1
        \\

        LiNbO$_3$ \cite{He2019selfstarting} &
        FWM, SHG&
        755-800 &    % pump is in between 1040-1065 nm
        1535-1545 &
        n.a.
        \\

        LiNbO$_3$ \cite{Roy2023VisiblemidIR} &
        \chitwo-sOPO, SHG, \chitwo-SFG &
        520--1200$^\mathrm{a}$ &    % pump is in between 1040-1065 nm
        1040--1065$^\mathrm{b}$ &
        n.a.
        \\

        LiNbO$_3$ \cite{Sekine2025multioctave} &
        \chitwo-sOPO, SHG, \chitwo-SFG &
        443--2676 &    
        1045$\pm$5$^\mathrm{b}$ &
        n.a.
        \\

        GaP \cite{Wilson2020Integratedgalliumphosphide} &
        SHG, FWM, \chitwo-SFG &
        765-795 &    
        1560 &
        n.a.
        \\

        4H-SiC \cite{Wang2022solitonformation} &
        SHG, FWM, \chitwo-SFG &
        770-800 &    
        1563 &
        n.a.
        \\

        AlN \cite{Jung2014GreenredIR} &
        FWM, SHG, \chitwo-SFG, \chithree-SFG &
        517, 776 &    
        1550, 1552 &
        n.a. %<$10^{-7}$ % off-chip power: >1.2 $\times 10^-6$ for red, >75 $\times 10^-9$ for green
        \\

        AlN \cite{Guo2018efficient} &
        SHG, FWM &
        720-820&    
        1550 &
        0.61 % off-chip power: >1.2 $\times 10^-6$ for red, >75 $\times 10^-9$ for green
        \\

        AlN \cite{Liu2018Generationmultiplevisible} &
        SHG, FWM, \chitwo-SFG&
        720-840&    
        1558.1 &
        $<1\times10^{-3}$ 
        \\

        AlN/\SiN\ \cite{Surya2018Efficientthirdharmonic} &
        THG&
        514&    
        1542 &
        $4.9\times10^{-2}$ 
        \\

        Ta$_2$O$_5$ \cite{Brodnik2025Nanopatternedparametricoscillators} &
        \chithree-sOPO &
        (749-897)/(1303-1806)$^\mathrm{a}$&  % 1303 from fig. 2f, 1/(2/1062-1/1303)
        $\sim$1062 &
        n.a.
        \\

        \SiN\ \cite{Miller2014} &
        FWM, SHG, \chitwo-SFG &
        765-775 &
        1540 & % estimated from their experiment pump wavelengths
        1
        \\

        \SiN\ \cite{Wang2016Frequencycombgeneration} &
        FWM, THG, \chithree-SFG &
        502-600 &
        1561-1565 & % estimated from their experiment pump wavelengths
        0.123
        \\

        \SiN\ \cite{CoratoZanarella2025Simultaneouschipgeneration} &
        \chithree-SFG, FWM &
        434-680 &
        1541-1571 & 
        n.a.
        \\

        \SiN\ \cite{Lu2019Milliwattthresholdvisible} &
        \chithree-sOPO &
        (714-781)/(1159-1484)$^\mathrm{a, c}$ &  %714.6-1316.9, 781.4-1484, 771.6-1187, 771.8-1159.2 &
        926.5-932.5, 1020-1024 &              % estimated from their experiment pump wavelengths
        <0.1/<0.01$^\mathrm{d}$               % the typical OPO signal and idler are 10 dB to 20 dB lower than the pump
        \\

        \SiN\ \cite{Lu2020chipopticalparametric} &
        \chithree-sOPO &
        (568-760)/(793-1181)$^\mathrm{a,c}$ &
        766-778 & % estimated from their experiment pump wavelengths
        <0.01/<0.01$^\mathrm{d}$
        \\

        \SiN\ \cite{Domeneguetti2021Parametricsidebandgeneration} &
        \chithree-sOPO &
        546.2/1465.3$^\mathrm{c}$, 613/1128$^\mathrm{c}$ &
        793-780 & % estimated from their experiment pump wavelengths
        <0.1/2$^\mathrm{d}$ 
        \\

        \SiN\ \cite{Stone2022Efficientchipbased} &
        \chithree-sOPO &
        (590-721)/(866-1150)$^\mathrm{a, c}$ &
        776-793 & % estimated from their experiment pump wavelengths
        5/4$^\mathrm{d}$
        \\

        \SiN\ \cite{Sun2024AdvancingchipKerr} &
        \chithree-sOPO &
        (532-633)/(1019-1447)$^\mathrm{a, c}$ & % 1019=1/(2/781-1/633)
        765-781 & 
        0.5/n.a.
        \\

        \SiN\ \cite{Long2024SubDopplerspectroscopy} &
        \chithree-sOPO &
        589/1150, 665/938, 703/877$^\mathrm{a}$ & 
        780 & 
        0.1/0.03
        \\

        \SiN\ \cite{Zhou2022HybridModeFamily} &
        hOPO &
        (714-731)/(810-837)$^\mathrm{c}$ & % estimated from fig 3(b)
        764-776 & % estimated from their experiment pump wavelengths
        4.6/5$^\mathrm{d}$
        \\

        \SiN\ \cite{Perez2023HighperformanceKerr} &
        hOPO &
        776.7/1103.4$^\mathrm{c}$ &
        973.4 & % estimated from their experiment pump wavelengths
        >1/>20$^\mathrm{d}$ % signal power estimated from spectra
        \\

        \SiN\ \cite{Yuan2025Efficientwavelengthtunable} &
        SHG &
        532 &
        1064&
        5.3
        \\

        \SiN\ \cite{wang2025integrated} &
        SHG, \chitwo-SFG &
        511-540 &
        1030-1070 &
        3.5
        \\

        \rowcolor{mygray}
        % \hline
        \makecell[c]{\textbf{\SiN} \\ \textbf{(this work)}} &
        \makecell[c]{\wtwow \textbf{hOPO, HG, \chitwo-SFG,}\\ \textbf{\chithree-SFG, sFWM}} &
        \makecell[c]{\textbf{UV ($\sim$397), visible (452} \\ \textbf{-483, 512--530, 575--600,} \\ \textbf{768-795), NIR (815--1480)}} &
        \makecell[c]{\textbf{1535--1590}} &
        \makecell[c]{\textbf{>10 for red,} \\ \textbf{>1 for green, yellow} \\ \textbf{orange, and NIR}} \\
        \hline
      \end{tabular}

      \begin{tablenotes}
        \footnotesize
        \item[] The table summarizes light generation at wavelengths shorter than the telecom band. sOPO, single-mode-family OPO. n.a.: data not available. 
        $^{\text{a}}$Obtained from multiple devices.
        $^{\text{b}}$Pulsed pump.
        $^{\text{c}}$Signal/idler wavelengths.
        $^{\text{d}}$Signal/idler power.
      \end{tablenotes}
    \end{center}
  \end{threeparttable}
\end{table*}

%\vspace{0.2cm}
\section*{Discussion and outlook}

\noindent We present the first demonstration of two octave-spanning light generation from UV to NIR range in a high-Q \SiN\ microring resonator, pumped by a monochromatic CW laser in the telecom band.
Initiated by the efficient AOP-enabled SHG that triggers multimode nonlinear coupling, diverse cascaded nonlinear processes, such as \wtwow hOPO, HG, SFG, and sFWM, generate milliwatt-level NIR, green, yellow, and orange light. This wide spectral coverage is enabled by the dynamic reconfiguration of the SH mode, which provides a critical control parameter for the phase-matching condition of \wtwow hOPO and the resulting cascaded processes, exploiting the extra spatial degree of freedom for phase-matching and yielding highly tunable output. 

Our system stands out for its design simplicity, wide tunability, along with decent output power.
We emphasize that these properties are unique to \SiN\ photonic platform, which features highly reconfigurable, photo-induced \chitwo\ nonlinearity, as well as excellent \chithree\ response. 
Such performance would not be achievable in intrinsic \chitwo\ platforms, where the efficient SHG comes at the expense of flexibility, due to the requirement for static electric-field poling and/or perfect phase-matching. 
In terms of spectral coverage, while thin film lithium niobate (TFLN) outperforms other platforms when considering  supercontinuum generation in straight waveguides \cite{Wu2024Visibleultravioletfrequency, Ludwig2024Ultravioletastronomicalspectrograph} or broadband frequency combs in synchronously pumped OPOs \cite{Roy2023VisiblemidIR, Sekine2025multioctave}, these required pulsed pumps and bypassing noisy nonlinear states to preserve coherence. 
For a comprehensive comparison, Table \ref{table1} summarizes the state-of-the-art of nonlinear frequency conversion in planar microresonator systems from several integrated photonic platforms.
Enabled by the newly discovered \wtwow hOPO process, our results demonstrate remarkable wavelength access and output power compared to previous realizations, albeit with a relatively high pump power requirement, which could be reduced by improving SHG conversion efficiency, for example, in higher finesse microresonators \cite{Lu2021efficientphotoinduced}.

While all reported results in this work are obtained from a single device, we anticipate even denser and/or broader spectral coverage through on-chip integration of multiple microresonators with controlled geometry variations, as well as higher nonlinear conversion efficiency in the visible range through dedicated coupler designs \cite{Lu2021efficientphotoinduced, Yuan2025Efficientwavelengthtunable}.
The same design principle of \wtwow hOPO could also be implemented at shorter wavelengths, allowing efficient generation of visible signal and idler fields regardless of the strong normal dispersion. Tailored dispersion engineering of \wtwow hOPO process, could also promise a path towards mid-infrared light generation for full-spectrum on-chip sources.

From a fundamental perspective, the intriguing \wtwow hOPO dynamics that we demonstrate provides a foundational framework for harnessing cooperative second- and third-order nonlinear optical processes, yielding several important implications.
For example, the \chitwo-\chithree\ based nonlinear conversion between telecom and visible/NIR wavelengths provides broader wavelength access than the constituent processes alone.
This capability not only interfaces scalable photonic chips with diverse atomic systems \cite{Lu2024Emergingintegratedlaser} for time and frequency metrology, but could also enable telecom-to-visible quantum frequency conversion \cite{Li2016Efficientlownoise, Lu2019Efficienttelecomvisible, Wang2022Syntheticfivewave, Li2025downconverted, Raghunathan2025Telecomvisiblequantum} for long-distance quantum communications.

\newpage
\section*{Methods} \label{method}

\vspace{0.2cm}
\noindent{\textbf{Experimental optical setup.}} 
The light from an external-cavity diode lasers (ECDL) tunable in the C- and L-bands is amplified by an Erbium-doped fiber amplifier (EDFA) in the corresponding band. The light is adjusted to transverse-electric (TE) polarization using a fiber polarization controller (FPC) and coupled to the chip via a lensed fiber. An aspheric lens collects the light emitted from the chip output. The transmitted pump is reflected by a dichroic mirror (DM) to an IR photodetector (PD). The following DMs are removable, allowing the remaining light to be routed either to the signal and SH PDs, or directly to an optical spectrum analyzer (OSA).

\vspace{0.2cm}
\noindent{\textbf{Coupling losses and conversion efficiency estimation.}} 
For the calculation of conversion efficiency of the SHG and \wtwow OPO processes, the on chip power for the pump and the SH have been considered. 
Coupling losses have been measured at 1590~nm and 780~nm. We estimate approximately 2.7~dB and 3.5~dB losses per facet for the fundamental TE00 modes at 1550~nm and 780~nm, respectively. 

In Fig. \ref{fig2}b, a total loss of 28.5 dB is estimated for coupling from the chip to the OSA for the TE40 SH , based on the TE00 facet coupling loss.
Since the collection coupling is optimized for SH modes, the signal/idler are underestimated in the OSA peaks when using the value of TE00 SH coupling loss as a reference.

\vspace{0.2cm}
\noindent{\textbf{Data availability}} \\
\noindent The data and code that support the plots within this paper and other findings of this study are available from the corresponding author upon reasonable request.

\vspace{0.2cm}
\noindent{\textbf{Acknowledgements}}\\
\noindent This work has received funding from the Swiss State Secretariat for Education, Research and Innovation (SERI) number REF-1131-52105 (SERI- funded ERC AdG CHAGALL) and by the Swiss National Science Foundation (SNSF grant MINT 214889).

\vspace{0.2cm}
\noindent{\textbf{Author Contributions}}\\
\noindent S.S. and C.-S.B. conceived the project. S.S., J.Z., and B.Z. designed and performed the experiments with the help from M.C., O.Y., and C.L. S.S. and J.Z. developed the theoretical analysis with assistance from O.Y. The data analysis were carried out by S.S. and J.Z. S.S., J.Z. and C.-S.B. wrote the manuscript with input from M.C. and C.L. The \SiN \ samples were fabricated by J.L. under the supervision of T.J.K. The project was supervised by C.-S.B.

\vspace{0.2cm}
\noindent{\textbf{Competing interests}}\\
\noindent The authors declare no competing interests.

% \nolinenumbers
\bibliography{refs.bib}

\end{document}

\title{Supplementary Information for: Reconfigurable Resonant Multimode Nonlinear Coupling for UV-to-infrared Frequency Generation}

\author{Samantha Sbarra\orcidlink{0000-0002-0893-927X}}
\thanks{These authors contributed equally to this work.}
\affiliation{Photonic Systems Laboratory (PHOSL), STI-IEM, {\'E}cole Polytechnique F{\'e}d{\'e}rale de Lausanne,  CH-1015 Lausanne, Switzerland}

\author{Ji Zhou\orcidlink{0000-0001-8044-4426}}
\thanks{These authors contributed equally to this work.}
\affiliation{Photonic Systems Laboratory (PHOSL), STI-IEM, {\'E}cole Polytechnique F{\'e}d{\'e}rale de Lausanne, CH-1015 Lausanne, Switzerland}

\author{Boris Zabelich\orcidlink{0000-0002-8542-7542}}
\affiliation{Photonic Systems Laboratory (PHOSL), STI-IEM, {\'E}cole Polytechnique F{\'e}d{\'e}rale de Lausanne,  CH-1015 Lausanne, Switzerland}

\author{Marco Clementi\orcidlink{0000-0003-4034-4337}}
\affiliation{Photonic Systems Laboratory (PHOSL), STI-IEM, {\'E}cole Polytechnique F{\'e}d{\'e}rale de Lausanne,  CH-1015 Lausanne, Switzerland}
\affiliation{Present address: Dipartimento di Fisica “A. Volta”, Università di Pavia, 27100 Pavia, Italy}

\author{Christian Lafforgue\orcidlink{0009-0003-2027-9168}}
\affiliation{Photonic Systems Laboratory (PHOSL), STI-IEM, {\'E}cole Polytechnique F{\'e}d{\'e}rale de Lausanne,  CH-1015 Lausanne, Switzerland}

\author{Ozan Yakar\orcidlink{0000-0003-1357-8920}}
\affiliation{Photonic Systems Laboratory (PHOSL), STI-IEM, {\'E}cole Polytechnique F{\'e}d{\'e}rale de Lausanne,  CH-1015 Lausanne, Switzerland}

\author{Junqiu Liu\orcidlink{0000-0003-2405-6028}}
\affiliation{Laboratory of Photonics and Quantum Measurements (LPQM), SB-IPHYS, {\'E}cole Polytechnique F{\'e}d{\'e}rale de Lausanne,  CH-1015 Lausanne, Switzerland}
\affiliation{Present address: International Quantum Academy and Shenzhen Futian SUSTech Institute for Quantum Technology and Engineering, Shenzhen 518048, China}
\affiliation{Present address: Hefei National Laboratory, University of Science and Technology of China, Hefei 230088, China}

\author{Tobias J. Kippenberg\orcidlink{0000-0002-3408-886X}}
\affiliation{Laboratory of Photonics and Quantum Measurements (LPQM), SB-IPHYS, {\'E}cole Polytechnique F{\'e}d{\'e}rale de Lausanne,  CH-1015 Lausanne, Switzerland}

\author{Camille-Sophie Br\`es\orcidlink{0000-0003-2804-1675}}
\email{\textcolor{magenta}{camille.bres@epfl.ch}}
\affiliation{Photonic Systems Laboratory (PHOSL), STI-IEM, {\'E}cole Polytechnique F{\'e}d{\'e}rale de Lausanne,  CH-1015 Lausanne, Switzerland}

\maketitle

\tableofcontents

\setcounter{figure}{0}

\newpage
\subsection*{Supplementary Note I: Nonlinear pathways for visible and UV light generation}
\begin{figure*}[h]
	% \centering
	\includegraphics[width=1\linewidth]{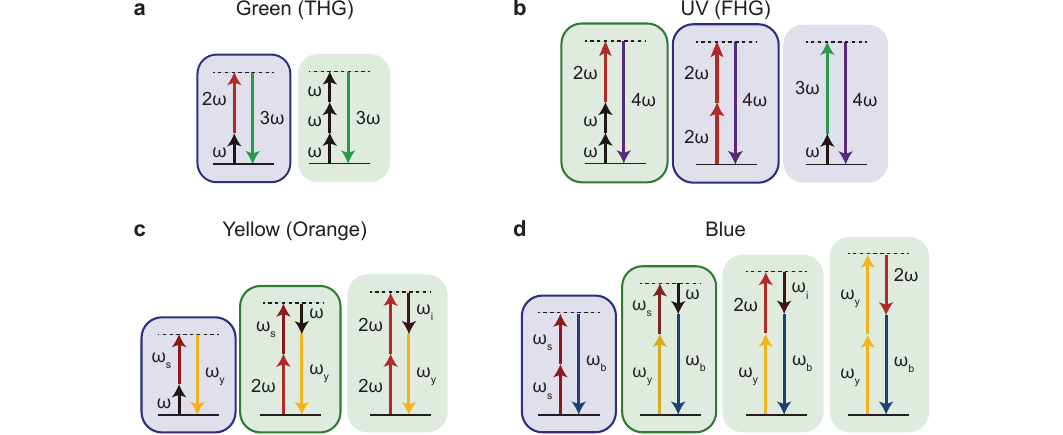}
	\caption{Energy-level diagrams of the nonlinear processes for the generation of light in the visible and UV spectral regions:
    \textbf{a}, third harmonic (green);
    \textbf{b}, fourth harmonic (UV);
    \textbf{c}, yellow–orange;
    \textbf{d}, blue.
    The background colors distinguishe three-wave mixing (TWM, blue) from four-wave mixing (FWM, green) interactions, while the borders highlight the processes already shown in Fig.~1c of the main text. The frequencies $\omega_{\mathrm{s}}, \omega_{\mathrm{i}},  \omega_{\mathrm{y}}$, and $\omega_{\mathrm{b}}$ correspond to the signal, idler, yellow, and blue spectral components, respectively.
	}
\label{figs1}
\end{figure*}

\noindent In Fig.~1c of the main text, we illustrated a subset of the nonlinear processes potentially responsible for the generation of spectral components in the visible and UV regions. There, we focused on those involving the fundamental telecom pump and \chitwo\ processes, as they are expected to be the most efficient pathways. However, additional mechanisms may also contribute.
For completeness, in Fig.~\ref{figs1} we summarizes all the identified processes that could account for the observed (a) green from third harmonic generation (THG), (b) UV from fourth harmonic generation (FHG), (c) yellow (orange), and (d) blue. 

\subsection*{Supplementary Note II: Characterization of the microresonator}

\begin{figure*}[b]
	\centering
	\includegraphics[width=1\linewidth]{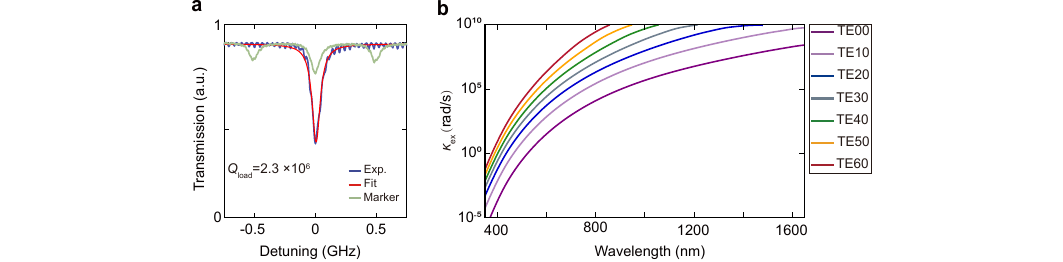}
	\caption{Linear characterization of the microresonator.
    \textbf{a}, Linewidth and Q factor measurement of the resonance at 1587.5 nm;
    \textbf{b}, Simulated dependence of the coupling rate $\Gamma$ on wavelength and mode index.}
\label{figs2}
\end{figure*}

\noindent In addition to the coupling losses measurement detailed in the Methods, we characterized the linewidth measurement of the representative resonance at around 1587.5 nm studied in the main text, with the results shown in Fig.~\ref{figs2}a. The optical transmission data (blue line) was measured by scanning a continuous-wave (CW) laser across the resonance and fitted with a Lorentzian lineshape (red line). For frequency calibration, the laser was modulated with a 0.5 GHz tone, generating sidebands that appear as two additional resonances (green line) and serve as frequency markers. From the calibrated linewidth (81.4 MHz), we extract the intrinsic and loaded Q factors of $13.7\times10^6$ and $2.3\times10^6$, respectively. 

Fig.~\ref{figs2}b shows the simulated coupling rates $\kappa_{\mathrm{ex}}$ for all considered TE modes. For the identified modes of the hOPO pump/SH/signal/idler pairs, the coupling rates were combined with the mode overlap integral $\Gamma$, to simulate the \wtwow hOPO spectra shown in Fig. 4a in the main text. $\Gamma$ is defined as \cite{Lin2008proposalhighlytunable}:
\begin{equation}
    \Gamma \equiv \frac{\iint{F_{i}^{*}F_jF_{k}^{*}F_l\,\mathrm{d}r\,\mathrm{d}z}}{\left( \iint{|F_i|^4\,\mathrm{d}r\,\mathrm{d}z} \right) ^{1/4}\left( \iint{|F_j|^4\,\mathrm{d}r\,\mathrm{d}z} \right) ^{1/4}\left( \iint{|F_k|^4\,\mathrm{d}r\,\mathrm{d}z} \right) ^{1/4}\left( \iint{|F_l|^4\,\mathrm{d}r\,\mathrm{d}z} \right) ^{1/4}}
\end{equation}
where $F_{i,j,k,l}$ are the TE transverse mode profiles at $\omega_{i,j,k,l}$. $r$ and $z$ denote the radial and axial coordinates, respectively.

\begin{figure*}[t!]
	\centering
	\includegraphics[width=1\linewidth]{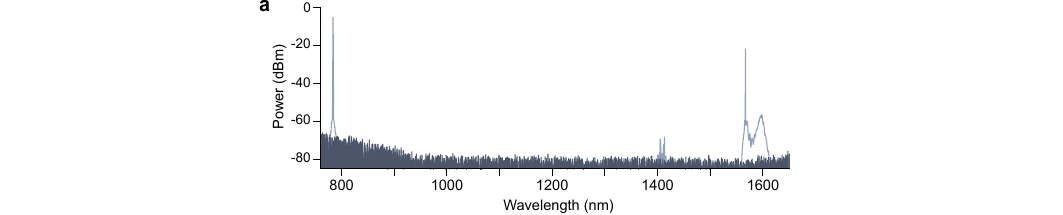}
	\caption{\textbf{a}, Pure SHG spectrum for estimation of the maximum SHG conversion efficiency.}
\label{figs2-1}
\end{figure*}

Fig.~\ref{figs2-1} shows the pure SHG spectrum used to estimate the maximum SHG conversion efficiency. With an SH peak at -5.2 dBm and a chip-to-OSA coupling loss of 28.5 dB, we estimate the maximum conversion efficiency $\eta_{\mathrm{SHG}}$ to be approximately 18\%/W with a pump power of about 1 W.

\subsection*{Supplementary Note III: Simulation of \chitwo\ grating periods for different SH mode orders}

\noindent Experimentally, we use two-photon microscopy (TPM) to measure the period of the optically inscribed $\chi^{(2)}$ grating, enabling identification of the generated second harmonic (SH) mode \cite{Nitiss2022Opticallyreconfigurablequasi}. We considers a fixed pump in the fundamental TE00 mode at 1587.5~nm, while the SH mode ranges from the fundamental TE up to the seventh order. From the quasi-phase-matching (QPM) condition, the grating period $\Lambda$ is defined by:
\begin{equation}
    \Lambda = \frac{\lambda_{\omega}}{2|n_{\mathrm{eff}}^{2\omega}-n_{\mathrm{eff}}^{\omega}|}
\end{equation}
where $n_{\mathrm{eff}}^{\omega}$ and $n_{\mathrm{eff}}^{2\omega}$ are the effective refractive indices of pump and SH, respectively. Table \ref{tables1} lists all the simulated azimuthal periods ($\Lambda$) corresponding to the excitation of different second harmonic (SH) mode orders. Note that the TPM image corresponds to the intensity pattern of the grating ($|\chi^{(2)}|^2$), the actual grating period is twice the period measured from the TPM image.

\begin{table}[htbp]
	\centering
	\caption{Simulated $\chi^{(2)}$ grating periods for different SH mode orders, when pumped at 1587.5 nm.}
	\label{tables1}
	\renewcommand{\arraystretch}{1.25}
	\begin{tabular}{|p{2cm}<{\centering}|p{2cm}<{\centering}|p{4cm}<{\centering}|}
		\hline
		Pump & SH & grating period $\Lambda$ (µm) \\ \hline
		\multirow{7}{*}{TE00} & TE00 & 4.08 \\ \cline{2-3} 
		& TE10 & 4.49 \\ \cline{2-3} 
		& TE20 & 5.39 \\ \cline{2-3} 
		& TE30 & 7.52 \\ \cline{2-3} 
		& TE40 & 15.54 \\ \cline{2-3} 
		& TE50 & 49.05 \\ \cline{2-3} 
		& TE60 & 8.31 \\ \hline
	\end{tabular}
\end{table}

\newpage
\subsection*{Supplementary Note IV: Theoretical identification of signal/idler frequencies and their modes}

\begin{figure*}[h]
	\centering
	\includegraphics[width=1\linewidth]{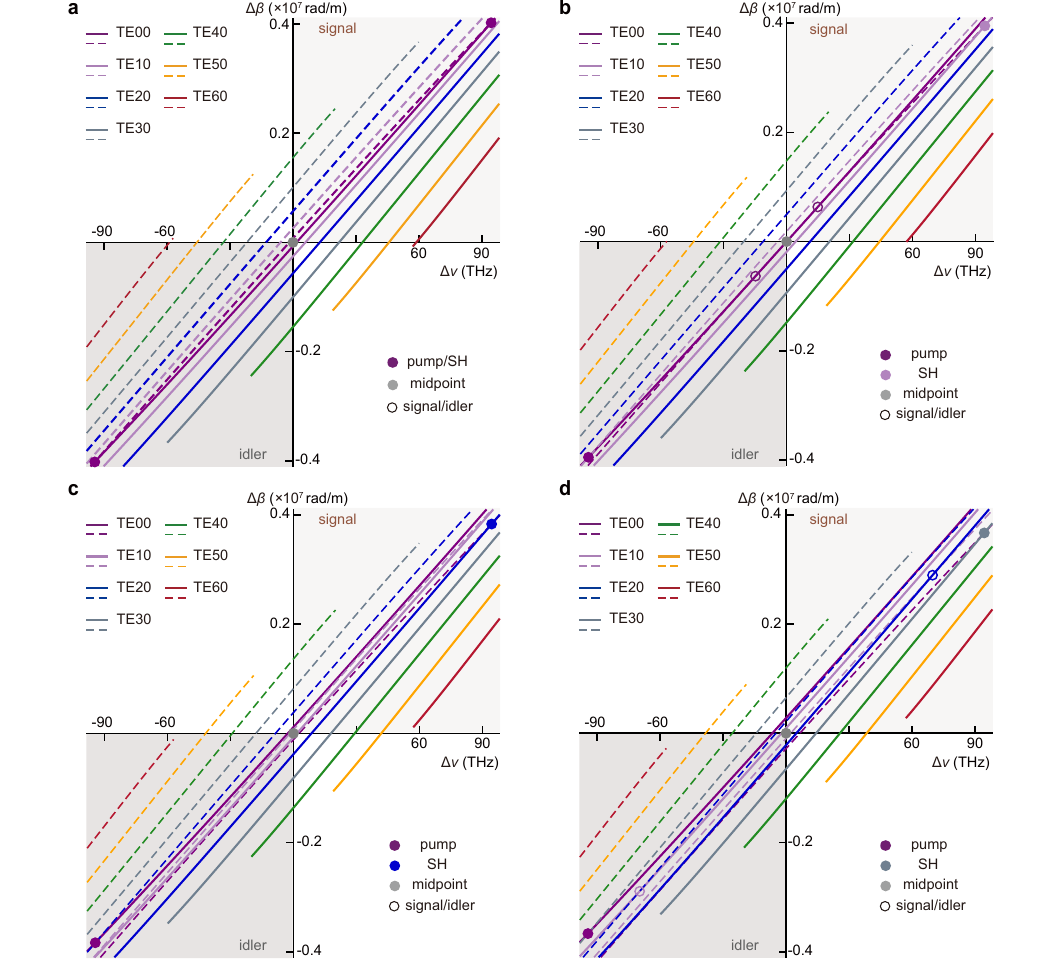}
        % \caption*{}
        \caption{Simulation of \wtwow hOPO signal and idler frequencies with a fixed TE00 pump at 1587.50 nm and a tunable SH in varying modes. The SH mode is switched between different transverse modes: \textbf{a)} TE00, \textbf{b)} TE10, \textbf{c)} TE20, \textbf{d)} TE30, \textbf{e)} TE40, \textbf{f)} TE50, and \textbf{g)} TE60. SH modes TE00 and TE20 do not support \wtwow hOPO process.}
        \label{figs3}
\end{figure*}

\noindent In Fig. 3c of the main text, for readability we present only the identified pump/SH/signal/idler \wtwow hOPO combination at modes TE00/TE40/TE30/TE10 with pump wavelength of 1587.50 nm. For a comprehensive overview, Fig. \ref{figs3} shows the simulated results for all considered SH modes with a TE00 pump at the fixed wavelength. Experimentally, since the resonances of different SH modes are inherently offset from each other and the doubly resonant condition for each depends exclusively on pump thermal tuning, we are able to excite only a few of these SH modes at a given pump wavelength. Nonetheless, the \chitwo\ nonlinearity in \SiN\ offers superior optical reconfigurability, tunability, and experimental simplicity compared to other material platforms.

\begin{figure*}[t!]
	\centering
	\includegraphics[width=1\linewidth]{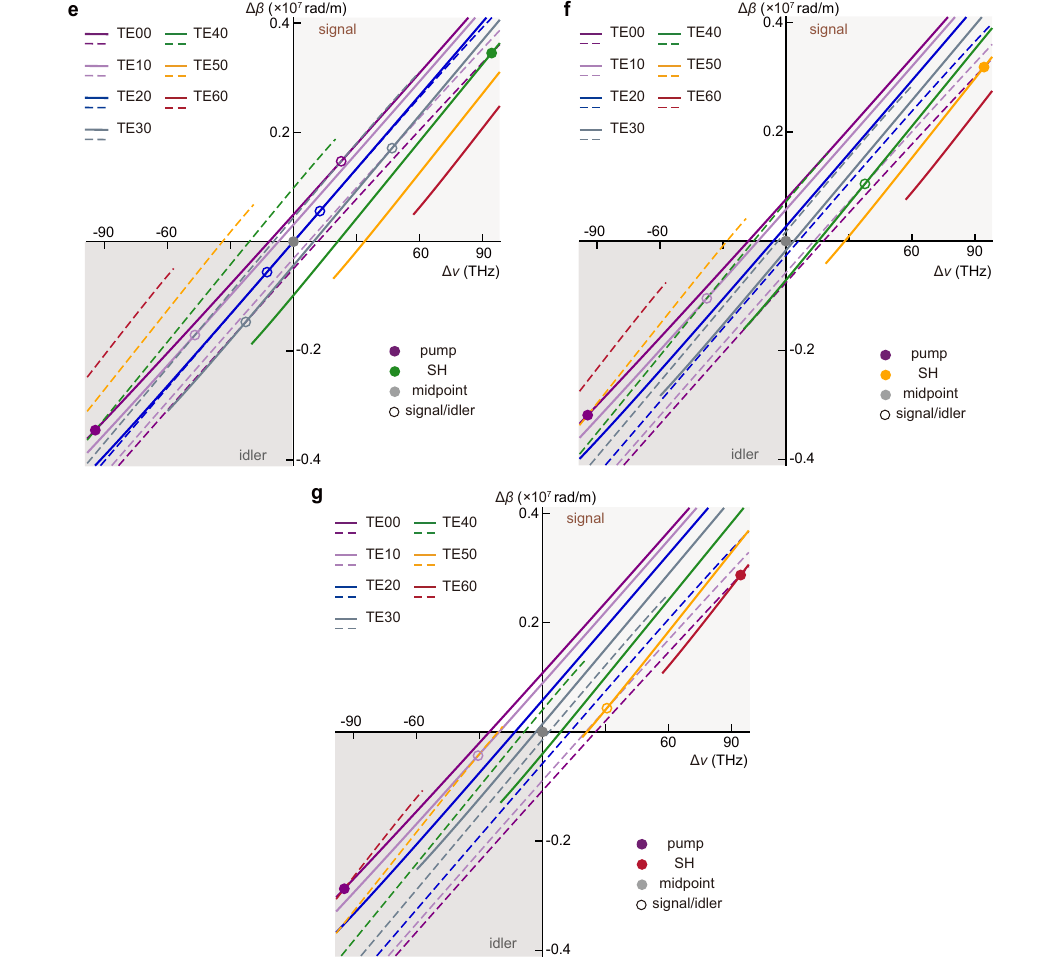}
        \caption*{}
\end{figure*}

\subsection*{Supplementary Note V: Other nonlinear phenomena}

\begin{figure*}[t!]
	\centering
	\includegraphics[width=1\linewidth]{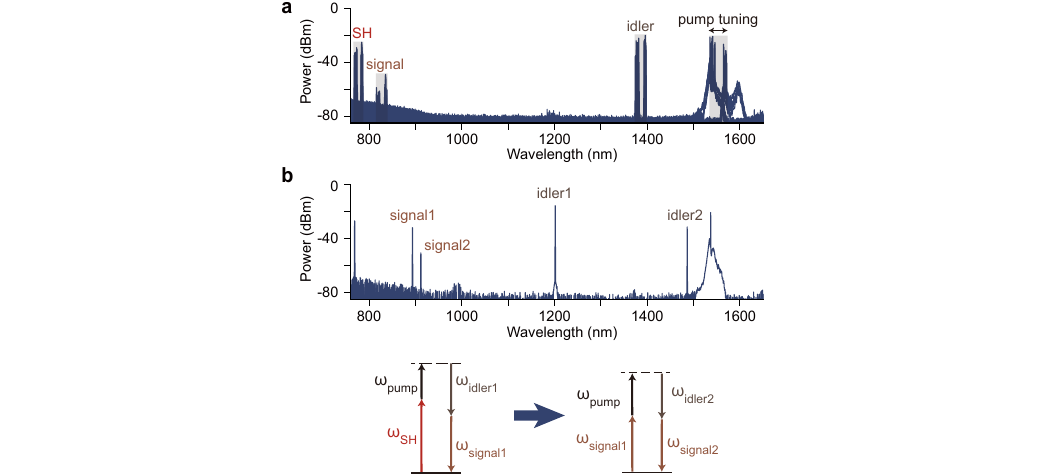}
        \caption{\textbf{a}, Validation of 1:1 output–input frequency tuning ratio in the \wtwow hOPO. The pump, SH, signal, and idler modes are TE00, TE30, TE20, and TE10, respectively. 
        \textbf{b}, Coexistence of two hOPO pairs for the pump/SH/signal/idler wavelengths at 1570.2/785.1/892.0/1266.6 nm and 1570.2/785.1/938.8/1182.9 nm, with the modes inferred as TE00/TE40/TE30/TE10 and TE00/TE60/TE50/TE10, respectively.}
\label{figs4}
\end{figure*}

\noindent In addition to the intriguing phenomena of hOPO coexistence and parametrically driven Kerr comb discussed in the main text, we present in Fig. \ref{figs4}a the experimental validation of fine wavelength tuning for the signal/idler output. From the simulation of Fig. 4a in the main text, the modes of pump/SH/signal/idler are confirmed as TE00/TE30/TE20/TE10, respectively. 

In Fig.~\ref{figs4}b, the original hOPO between the pump and SH that generates signal1 and idler1 triggers a cascade of hOPO involving the pump and signal1, generating signal2 and idler2. Such an hOPO cascade provides access to NIR wavelength at 1,480 nm, inaccessible from any individual \wtwow hOPO process. In future work, we anticipate that the hOPO cascade could be extended towards longer wavelengths or even the mid-infrared range.

The observed periodicity in hOPO1 pair (around 25 pm from I to III to IV) may result from dynamic switching among adjacent signal and idler resonances within the same mode families. This behavior is facilitated by the relaxed multiply resonant condition, which allows the large SH detuning to be compensated by substantial detuning of the signal and/or idler fields.

\newpage
\bibliography{refs.bib}